# Three Different Ways Synchronization Can Cause Contagion in Financial Markets

**Naji Massad [1] and Jørgen Vitting Andersen [1,2,*]**

1. Centre d'Economie de la Sorbonne, Université Paris 1 Pantheon-Sorbonne, Maison des Sciences Economiques, 106-112 Boulevard de l'Hôpital, 75647 Paris, CEDEX 13, France; najimassaad@hotmail.com
2. CNRS and Centre d'Economie de la Sorbonne, Université Paris 1 Pantheon-Sorbonne, Maison des Sciences Economiques, 106-112 Boulevard de l'Hôpital, 75647 Paris, CEDEX 13, France
* Correspondence: jorgen-vitting.andersen@univ-paris1.fr

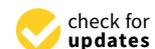

**Abstract:** We introduce tools to capture the dynamics of three different pathways, in which the synchronization of human decision-making could lead to turbulent periods and contagion phenomena in financial markets. The first pathway is caused when stock market indices, seen as a set of coupled integrate-and-fire oscillators, synchronize in frequency. The integrate-and-fire dynamics happens due to "change blindness", a trait in human decision-making where people have the tendency to ignore small changes, but take action when a large change happens. The second pathway happens due to feedback mechanisms between market performance and the use of certain (decoupled) trading strategies. The third pathway occurs through the effects of communication and its impact on human decision-making. A model is introduced in which financial market performance has an impact on decision-making through communication between people. Conversely, the sentiment created via communication has an impact on financial market performance. The methodologies used are: agent based modeling, models of integrate-and-fire oscillators, and communication models of human decision-making.

## 1. Introduction

Financial markets are generally thought of as random and noisy, beyond an understanding within an ordered framework. The elusive nature of the markets has been captured in theories like the efficient market hypothesis, which effectively treats price movements as random. It assumes that the price movements occurring on a given day constitute a random phenomenon, basically drawn from some probability distribution, thus describing in probabilistic terms what kind of event one should expect to happen on a given day. This assumption seems natural and probably has its roots back in the time when, at the beginning of a working day, people working in finance would turn on their radio and learn of new financial events (events assumed to be created by some higher authority). This kind of descriptive framework is also adopted for more modern and general models used in finance, such as autoregressive conditional heteroscedasticity (ARCH) and generalized autoregressive conditional heteroskedasticity (GARCH), which are able to describe many of the stylized facts observed in empirical data. Instead, socio-finance (Vitting Andersen and Nowak 2013) focuses on non-random human impacts on price formation in the financial markets, stressing in particular the interaction taking place between people, either directly through communication, or indirectly through the formation of asset prices, which will in turn be seen to enable synchronization in decision-making.

Synchronization in human decision-making, and the impact it could have on financial asset price formation is not a well understood topic. It is better known in economics, where empirical studies have shown that international trade partners display synchronization in business cycles. Dées and Zorell (2011) find that economic integration fosters business cycle synchronization across countries. A similar production structure is also found to enhance business cycle co-movement. On the other hand, these authors find it harder to pinpoint a direct relationship between bilateral financial linkages and output correlation. For other studies of synchronization and business cycles across countries, see, e.g., (Frankel and Rose 1998; Baxter and Kouparitsas 2005; Backus et al. 1992). The recent global financial crisis has raised questions about the role that financial market integration could have on synchronization of business cycles across borders (Rey 2015). However, very little research has been done on synchronization that is created endogenously by the financial markets themselves, without there necessarily being any economic cause. However, such phenomena could be relevant for both the onset and the continuation of a financial crisis (see, e.g., (Poledna et al. 2014; Aymanns and Georg 2015)). This naturally raises the following question: could synchronization endogenously created in financial markets spill over into the economy and thereby cause synchronization in business cycles across borders? It would seem here that it is important to have a clear framework for understanding such dynamics, and also the conditions for the onset of synchronization in decision-making.

It should be noted that the term "synchronization" in this article covers a broader phenomenon than "herding", a related term often used in the financial literature. In finance, "herding" usually refers to the simple case where people intentionally copy the behavior of others. It has been suggested that it is rational to herd (Devenow and Welch 1996). For instance, portfolio managers may mimic the actions of other portfolio managers just in order to preserve their reputation. It is easier to explain a failure when everybody else also fails than expose oneself to a failure due to bold forecasts and a departure from the consensus. For a general review paper on herding, see (Spyrou 2013).

Here, "synchronization" refers instead to the more general and complex case where people don't necessarily try to imitate each other's behavior, but, rather, by observing the same price behavior or through communication, end up synchronizing their decision-making. From this point of view, the synchronization described in this article may be closer to the idea of creation of convention put forward by Keynes (1936).

We note that one quantitative measure of synchronisation cannot possibly describe all aspects of syncrhonisation, especially since synchronisation can happen on different time scales, and between individuals, or a group of individuals. In the case with indirect interaction between individuals, we introduce the so-called decoupling parameter as a measure on synchronisation. In the case of direct interaction between group of market participants (e.g., between financial markets), we use as a measure for synchronisation how long a perturbation that begins in one market can propagate. In the case of direct interaction between agents, we use the tipping point analysis of sentiments, in order to identify when synchronisation is at its highest level.

Poledna et al. (2014) stress the way regulation policies could increase the amount of synchronized buying and selling needed to achieve deleveraging, which could, in turn, destabilize the market. They discuss the new regulatory measures proposed to suppress such behavior, but it is not clear whether these measures really address the problem. In addition, they show how none of these policies would be optimal for everyone: risk-neutral investors would prefer the unregulated case with low maximum leverage, banks would prefer the perfect hedging policy, and fund managers would prefer the unregulated case with high maximum leverage. Aymanns and Georg (2015) consider instead the case where banks choose similar investment strategies, which can in turn make the financial system more vulnerable to common shocks. They consider a simple financial system in which banks decide about their investment strategy on the basis of a private belief about the state of the world, and a social belief formed by observing the actions of peers. They show how the probability that banks will synchronize their investment strategies depends on the relative weighting of private and social belief.

In the following, we will place the emphasis on the fact that price formation is the result of human decisions to buy or sell assets. Behind every trade is a human decision, if not through the direct action of a human, then indirectly through programs that manages algorithmic trading by computers. Socio-finance (Vitting Andersen and Nowak 2013) treats price formation as a sociological phenomenon. It considers the dynamics of price formation to be created via either direct or indirect human interactions. Direct interaction covers the case where individuals or groups of individuals communicate directly and thereby influence mutual decision-making with respect to trading assets. At the first level, the individual level, indirect interaction covers the case where a trader submits an order to buy or sell an asset. The resulting price movement of the asset is observed by other traders, who may in turn modify their decision-making as a consequence of the price movement of the asset caused by the initial trade. At the second level, the group level, indirect interaction covers the case where whole markets await the outcome of pricing in other markets in order to find their own pricing level.

## 2. Three Different Ways Synchronization Can Lead to Contagion in Financial Markets

In the following, we will study how synchronization in decision-making can happen for people trading in financial markets. In the context of the present article, synchronization is defined as a certain dynamic in decision-making. Synchronization is defined as the dynamics where the decision of an individual, or groups of individuals, in turn influences the decision-making of other individuals or groups of individuals. Here, we are considering financial markets, so the decision-making is whether to buy an asset, to sell an asset, or to do nothing. Two different dynamics will be considered. Either the dynamics that happen because people communicate (direct interaction) or the dynamics that happen through and index (indirect interaction). The research question of the different mechanisms, and how they can be identified and studied are discussed in the following.

*2.1. Synchronization through Indirect Interaction of Traders*

This section is divided into two parts: indirect interaction of market participants at the individual level (Section 2.1.1) and indirect interaction of market participants at the collective level (Section 2.1.2). It should be noted that in this article we only treat indirect interaction of traders through a market index. Other situations where an indirect interaction would play an important role is, however, possible. The case where some important global news stories (e.g., terrorism on a global scale like 9/11) could also impact the decision-making of market participants without the need for an interaction between market participants. Such cases are, however, beyond the scope of the models introduced in the following.

2.1.1. Synchronization through Indirect Interaction of Individuals: The First Level

What traders have done in the past has a direct influence on the action of traders in the present. Past buying and selling activities bring the market up to the present level, in which traders have to decide whether now is an opportune moment to buy or sell. This applies to traders using technical analysis, as well as traders using fundamental analysis. Technical analysis will give different buy/sell signals, depending on the exact price history generated by traders in the past, whereas fundamental analysis will be used by traders to decide whether the price level has become low enough to buy or high enough to sell.

Thus, as traders take note of what happens in the market and update their trading strategies accordingly, this will change their future prospects for trade in the market. Therefore, as the markets evolve, traders' buying/selling decisions will change, and, as these decisions change, they will thereby modify the pricing of the market. This feedback loop is illustrated schematically in Figure 1. Note that here we assume that arbitrage possibilities are exploited instantaneously by the traders. This is in contrast to, for example, the study by the synchronization of Abreu and Brunnermeier (2002), where arbitrageurs become sequentially aware of mispricing. In that case, rational arbitrageurs instead

"time the market" rather than correct mispricing right away. In (Abreu and Brunnermeier 2002), this was shown to lead to delayed arbitrage, something which is not present in the models described in this study.

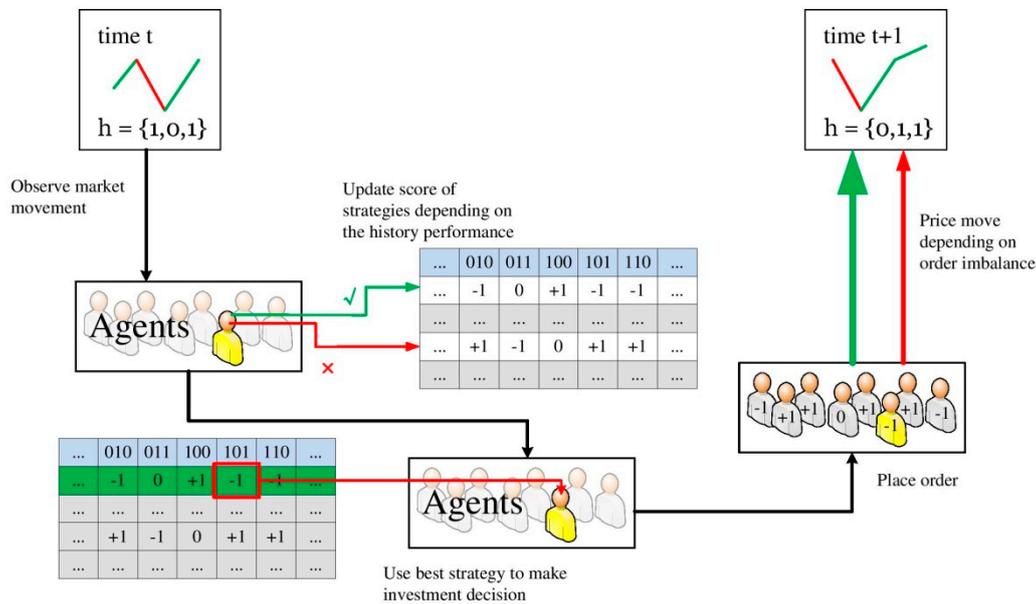

**Figure 1.** Representation of the price dynamics in the Minority Game (Challet and Zhang 1997) and the $-Game (Andersen and Sornette 2003). Agents first update the scores of all their strategies depending on their ability to predict market movements. After the scores have been updated, each agent chooses the strategy that now has the highest score. Depending on the price history at that moment, this strategy determines which action a given agent will take. Finally, the sum of the actions of all agents determines the next price move: if positive, the price moves up, if negative, the price moves down.

Under certain circumstances, it can happen that traders inadvertently end up in a state where their trading strategies "decouple" from the price history, so that, over the next (few) time step(s), their decision-making becomes completely deterministic, independent of what happens next in the market. In order to illustrate this point, consider the table below, which is one way of formalizing technical analysis trading strategies in a simple table form (Abreu and Brunnermeier 2002, 2003). Considering for simplicity only the direction of each of the last market moves, the table below predicts, for each possible price history, the next move of the market. The table illustrates one technical analysis strategy that uses the last three time periods to make a prediction and can easily be generalized to any number of periods.

Consider now a given market price history, $\vec{\mu}(t) = (010)$, at time $t$, which means that (as illustrated in Figure 2) three time periods ago the market went down, then up, and then down. It should now be noted that *whatever* the price movement in the next time period $t + 1$, the strategy in Table 1 will *always* predict to sell at time period $t + 2$. Therefore, we don't need to wait for the market outcome at the next time step $t + 1$ in order to know what the strategy will suggest in the subsequent time step: it will always suggest selling at time $t + 2$. The idea that such dynamics in the decision-making of technical analysis strategies could be relevant for real markets was suggested in (Andersen and Sornette 2003). In the terminology of (Andersen and Sornette 2003), the strategy in Table 1 is said to be "one time step decoupled, conditioned on the price history $\vec{\mu} = (010)$", and denoted $a_\mu^{decoupled}(t)$.

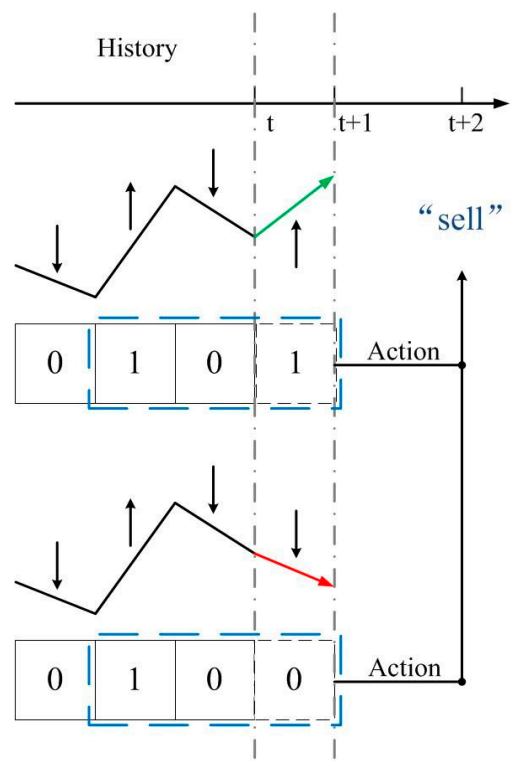

**Figure 2.** Representation of how the trading strategy in Table 1 decouples at time $t + 2$, conditioned on the price history $\vec{\mu} = (010)$ at time $t$.

**Table 1.** Example of a strategy used in the Minority Game (Challet and Zhang 1997) and the $-Game (Andersen and Sornette 2003). Considering only up (1) and down (0) market price movements, a strategy issues a prediction for each given price history, here illustrated with price histories over the last $m = 3$ days. A prediction of 1 refers to the recommendation of the strategy to buy, whereas a prediction of $-1$ is a recommendation to sell.

| Price History | Prediction |
|:---:|:---:|
| 0 0 0 | 1 |
| 0 0 1 | $-1$ |
| 0 1 0 | 1 |
| 0 1 1 | 1 |
| 1 0 0 | $-1$ |
| 1 0 1 | $-1$ |
| 1 1 0 | 1 |
| 1 1 1 | 1 |

We can then divide trading strategies into two different classes: those coupled to the price history (i.e., conditioned on knowing $\vec{\mu}(t)$, so that we cannot know the prediction of $a_\mu^{coupled}(t)$ at time $t + 2$ before knowing $\vec{\mu}(t+1)$), and those decoupled from the price history. Considering only the strategies actually used by agents to trade at time $t$, the order imbalance, $A(t)$, can therefore be written,

$$A(t) \equiv A_\mu^{coupled}(t) + A_\mu^{decoupled}(t), \tag{1}$$

where $A_\mu^{coupled}(t) = \sum a_\mu^{coupled}(t)$ is the sum over coupled strategies at time *t*, and similarly for $A_\mu^{decoupled}(t) = \sum a_\mu^{decoupled}(t)$. The condition for certain predictability at time *t*, two time steps ahead is then,

$$A_\mu^{decoupled}(t+2) > N/2. \qquad (2)$$

If a majority of market participants implement decoupled strategies, this will ensure a deterministic future price movement of the market, independently of the choices made by the minority that adopt coupled strategies.

Inequality (2) gives the condition for synchronization to happen via indirect interaction of traders through the price formation of an asset. Before considering synchronization in real markets (Challet and Zhang 1997), one must obviously begin by showing its presence in models, as well as in experiments. Figure 3 proves the existence of synchronization via decoupling in models like the Minority Game. This is a somewhat surprising result since, by definition, this type of game doesn't support trend-following strategies. For further references covering synchronization in experiments and markets, see (Challet and Zhang 1997; Andersen and Sornette 2003; Liu et al. 2016).

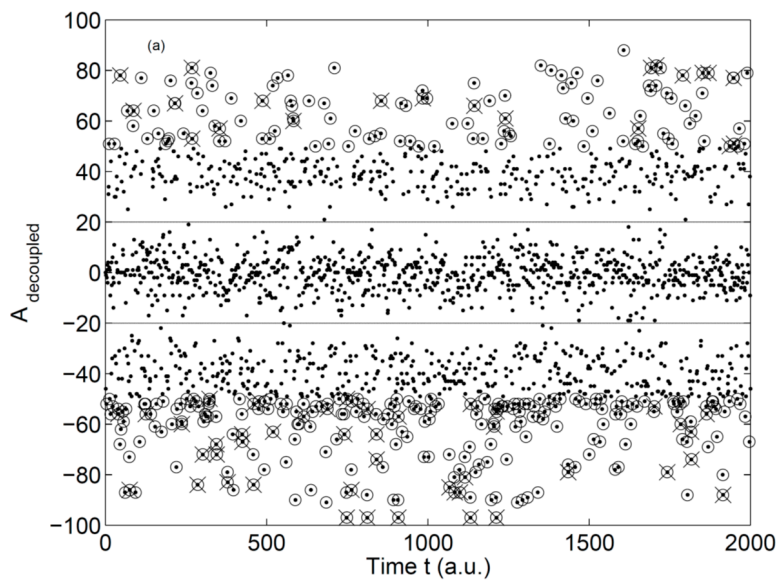

**Figure 3.** An example of $A_{decoupled}$ defined from (1) as a function of time for a simulation of the Minority Game. Circles indicate "one-step days" which are predictive with probability 1, while crosses indicate the subset of days starting a run of two or more consecutive one-step predictive days.

2.1.2. Synchronization through Indirect Interaction of Groups of Individuals: The Second Level

Having discussed how indirect interaction of individuals through financial indices can lead to synchronization, let us next consider how the phenomenon can appear through indirect interaction of groups of individuals. In this case, we consider the reaction of one market to the pricing created in another market. That is, we consider how a given market (i.e., a pool of traders) reacts to the previous price formation in another market (created by another pool of traders).

To illustrate this point, consider Figure 4, which shows how significant price movements of large capital stock indices can have a particular impact on smaller capital stock indices. The figure illustrates the effect of both a world market return (calculated as a weighted sum of returns of stock indices) and the US market return on the subsequent price movements of individual stock indices. Using the open–close return of the U.S. stock market, we see a particularly clear case of a "large-move" impact across markets: since the Asian markets close before the opening of the U.S. markets, they should only be able to price in this information when they open the next day. That is, one can consider the impact in the "close–open" of the Asian markets that follows *after* an "open–close" of the US market.

An eventual "large-move" U.S. open–close should therefore have a clear impact on the following close–open of the Asian markets. Figure 4 shows that this is indeed the case. In contrast, the European markets are still open when the U.S. market opens up in the morning, so the European markets have access to part of the history of the U.S. open–close. Any "large-move" U.S. open–close would therefore still be expected to have an impact on the following close–open of the European markets, but with larger variation in the response than for the Asian markets, since part of the U.S. move would already be priced in when the European markets closed. This is seen to be the case.

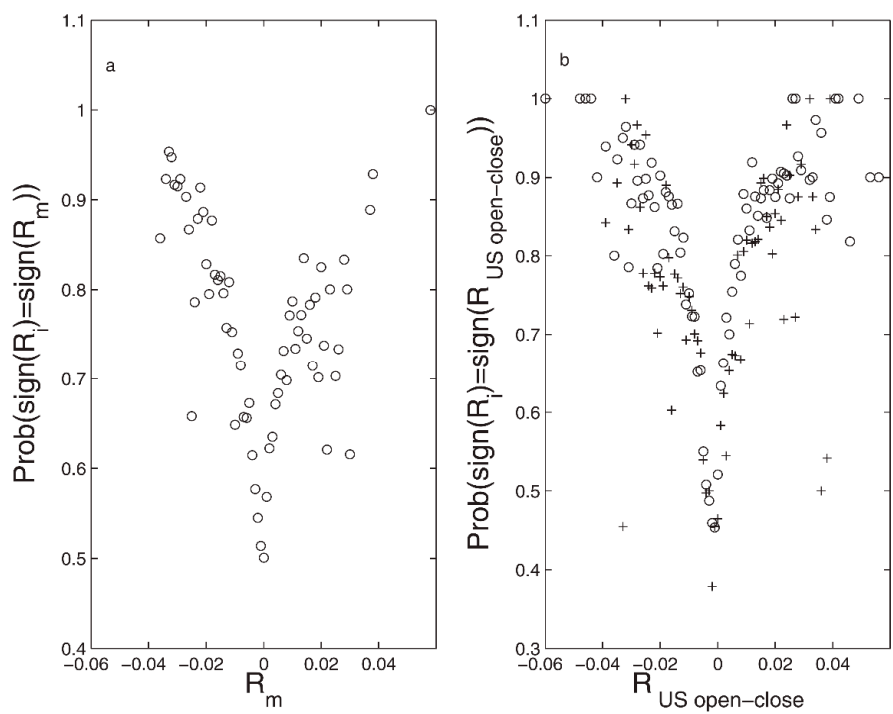

**Figure 4.** Illustration of change blindness: a large world market return (**a**) or US market return (**b**) impacts a given stock exchange, whereas small returns have random impact. (**a**) conditional probability that the daily return $R_i$ of a given country's stock market index has the same sign as the world market return; (**b**) conditional probability that the close–open return $R_i$ of a given country's stock market index following a U.S. open–close has the same sign as the U.S. open–close return (+: European markets; circles: Asian markets). The figures were created using almost nine years of daily returns from 24 of the major stock markets worldwide. For more information, see (Andersen and Sornette 2005; Liu et al. 2017).

To see how synchronization can happen across markets, consider the illustration in Figure 5a below, which shows three Integrate-And-Fire (IAF) oscillators with the same frequency over one time period (or equivalently one IAF oscillator over three time periods). An IAF oscillator is characterized by an accumulation (i.e., "integrate") in amplitude $A(t)$ (e.g., "stress") over time $t$, up to a certain point $A_C$, after which it discharges (i.e., "fires"). IAF oscillator models become complex when the oscillators are coupled (i.e., the amplitude of one oscillator influences the amplitude of other oscillators), and have different frequencies (see Figure 5b) and/or thresholds $A_C^i$. Peskin (Andersen et al. 2011) introduced IAF oscillators in neurobiology to describe interactions between neurons, but IAF oscillators have been introduced in many other contexts. For network studies of IAF oscillators see, e.g., Bellenzier et al. (2016); Kuramoto (1991); and Bottani (1995). A link between certain types of integrate-and-fire oscillators and earthquake models has also been noted by, e.g., Corral et al. (1995).

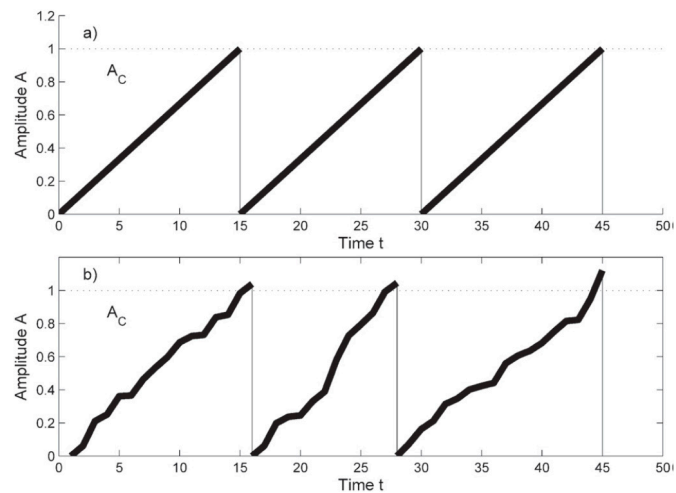

**Figure 5.** Illustration of an integrate and fire (IAF) oscillator. (**a**) the case where the amplitude *A*(*t*) of an IAF oscillator integrates linearly in time until it reaches a critical value Ac, after which it discharges by setting *A*(*t*) = 0. The case in (**a**) can be viewed as one IAF oscillating over three periods of time, or equivalently, three identical and uncoupled IAF oscillators oscillating over one period of time; (**b**) an IAF oscillator with random frequency over three time periods, or, equivalently, three different unit oscillators with random frequency over one time period. The figure is taken from (Liu et al. 2017).

As mentioned in (Andersen et al. 2011), one can consider each financial market index as an IAF oscillator that influences other market indices (i.e., other IAF oscillators). The impact, or "stress", from index *i* on index *j* accumulates up to a certain point, after which it becomes priced in. The justification for such a behavior can be seen from Figure 4, which shows that small price changes in index *i* have no immediate influence on index *j* (but are assumed to accumulate over time), whereas large price changes in index *i* have an impact and thereby become priced into index *j*.

This can be formalized in the following expression, which expresses the set of stock market indices worldwide as a set of coupled IAF oscillators:

$$R_I(t) = \frac{1}{N_i^*} \sum_{j \neq i}^{N} \alpha_{ij} \theta(|R_{ij}^{cum}(t-1)| > R_C) \times R_{ij}^{cum}(t-1)\beta_{ij} + \varphi_{ij}(t), \qquad (3)$$

$$R_{ij}^{cum}(t) = [1 - \theta(|R_{ij}^{cum}(t-1)| > R_C)] \times R_{ij}^{cum}(t-1) + R_j(t), \; j \neq i, \qquad (4)$$

$$\alpha_{ij} = 1 - \exp[-K_j/(K_i \, \gamma)]; \; \beta_{ij} = \exp[\frac{-|Z_i - Z_j|}{\tau}] \qquad (5)$$

In Label (3), $R_i(t)$ is the return of stock index *j*, which at time *t* receives a contribution from stock index *j*, whenever the "stress" $R_{ij}^{cum}$ exceeds a certain threshold $R_C$. $\alpha_{ij}$ describes the coupling between the two stock indices, expressed via Label (5) in terms of the relative weight of capitalizations $K_i$. A large $\gamma$, $\gamma \gg 1$, corresponds to a network of the world's indices dominated by the index with the largest capitalization Kmax, while a small $\gamma$, $\gamma \ll 1$, corresponds to a network of indices with equal strengths, since $\alpha_{ij}$ then becomes independent of *i* and *j*. In addition, it is assumed that countries that are geographically close also have greater economic interdependence, as described by the coefficient $\beta_{ij}$, where $|z_i - z_j|$ is the time zone difference between countries *i* and *j*. $\tau$ gives the scale over which this interdependence declines. Small $\tau$, $\tau \ll 1$, then corresponds to a world where only indices in the same time zone are relevant for pricing, whereas large $\tau$, $\tau \gg 1$ describes a global influence on the pricing that is independent of the difference in time zone.

It can be seen from Label (4) that it is the tensor $R_{ij}^{cum}$ that plays the role of an IAF oscillator. Returns from index *j*, $R_j$, accumulate stress on index *i* by continuously adding to $R_{ij}^{cum}$, up to a certain point, $\left|R_{ij}^{cum}\right| > R_C$, after which the oscillator discharges, $R_{ij}^{cum} \to 0$, and the stress becomes priced in via Label (3).

Once the IAF network is in a state of synchronization, contagion effects can be identified throughout the network. One example is given in Figure 6, showing the propagation of a large price movement taking place in the Japanese stock market on 23 May 2013. For more examples with real market data, see (Liu et al. 2017).

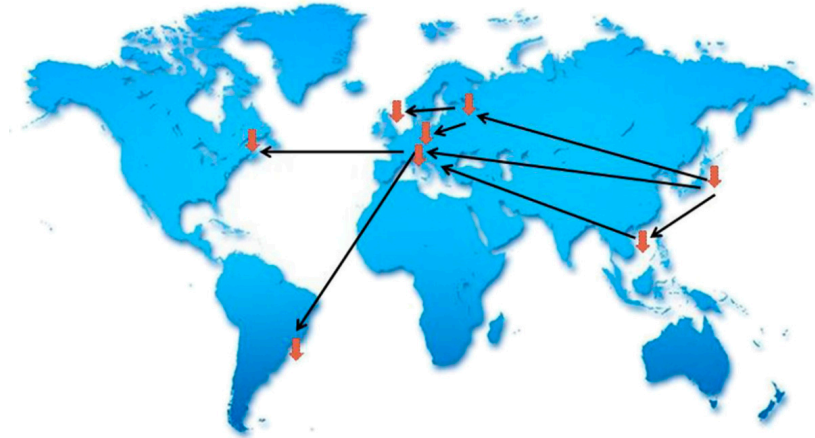

**Figure 6.** A price-quake. One of the main advantages of the nonlinearity in the integrate-and-fire oscillator model is that it enables a clear-cut identification of cause and effect. The figure shows one example of a price-quake, following an initial minus 7% price movement on the Japanese stock market on 23 May 2013. The figure is taken from (Liu et al. 2017).

*2.2. Synchronization through Direct Interaction of Traders*

Synchronization through Direct Interaction of Individuals and Groups of Individuals: The First and Second Levels

Having discussed how synchronization can emerge through indirect interaction of individuals, or groups of individuals, via financial market indices, we now consider the case of direct interaction, that is, ways that decision-making can be influenced by direct communication between people or groups of people. The idea is to see how discussions among market participants can influence their decision-making with respect to buying/selling assets, and how this can in turn influence market performance. We will also show how the market performance itself can be a relevant factor in the decision-making process, thereby creating another feedback loop between decision-making and market performance.

To see how this can happen, consider Figure 7A, which shows a population of market participants with different views of the market. For simplicity, we consider two possible states for these views: positive, bullish, or negative, bearish. Figure 7A shows an example where half the population is initially bearish, and the other half bullish. One could, for example, imagine a morning meeting taking place in a major bank or brokerage house, so, at the beginning of the day, we let people meet and discuss things in groups of different sizes (Figure 7B). To illustrate how communication between people can influence their decision-making, consider first the simple case where consensus decision-making is determined by the majority opinion (Figure 7C). As can be seen in Figure 7D, at the end of the day, the opinion of the population has changed as a result of their meetings (direct interaction), with only 45% of the population now bullish.

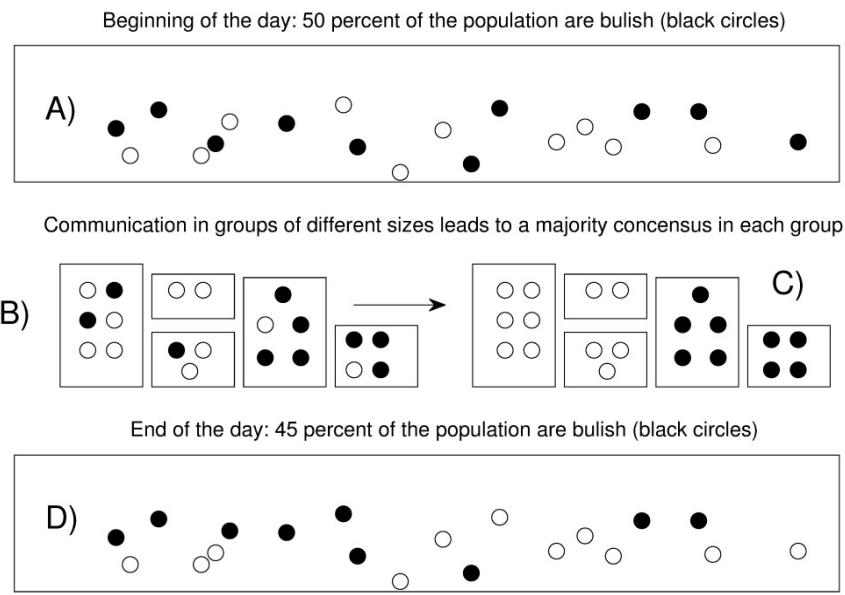

**Figure 7.** Changing the "bullishness" in a population via communication in subgroups. (**A**) at the beginning of a given day $t$, a certain percentage $B(t)$ are bullish; (**B**) during the day communication takes place in random subgroups of different sizes; (**C**) illustrates the extreme case of complete polarization $m_{k,j} = \pm 1$ created by a majority decision. In general, $m_{k,j} \simeq j/k$ corresponds to the neutral case where opinion remains unchanged on average within a subgroup of size $k$; (**D**) due to the communication in the different subgroups, the "bullishness" at the end of the day is different from what it was at the beginning of the day. The figure is taken from (Corral et al. 1995).

In the context of decision-making about trading assets in financial markets, it is natural to assume that market performance itself could influence the decision-making of market participants, while this could in turn influence future market performance. A model was suggested in (Corral et al. 1995) to capture this kind of feedback. The main idea is to let market performance influence decision making, rather than just going by a simple majority rule as illustrated in Figure 7B,C. This is done by assuming a certain *probability* for a majority opinion to prevail. Then, under certain conditions, a minority could persuade a part of the majority to change their opinion. The probability of a majority opinion prevailing will depend on the market performance over the last time period.

Specifically, let $B(t)$ denote the proportion of bullishness in a population at time $t$, whence the proportion of bearishness is $1 - B(t)$. For a given group of size $k$ with $j$ agents having a bullish opinion and $k - j$ a bearish opinion, we let $m_{k,j}$ denote the transition probability for all ($k$) members to adopt the bullish opinion as a result of their meeting. After one update taking into account communications in all groups of size $k$ with $j$ bullish agents, the new probability of finding an agent with a bullish view in the population can therefore be written

$$B(t+1) = m_{kj}(t) C_j^k B^j [1 - B(t)]^{k-j}, \qquad (6)$$

where

$$C_j^k \equiv \frac{k!}{j!(k-j)!}. \qquad (7)$$

It should be noted that the transition probabilities $m_{k,j}(t)$ depend on time, since we assume that they change as the market performance evolves.

The link between communication and its impact on the markets can now be taken into account by assuming that the price return $r(t)$ changes whenever there is a change in the bullishness. It should now be noted that it is the *changes* in opinion that matter for the market performance, rather than the *level* of a given opinion. Empirical data supporting this idea can be found in (Andersen et al. 2014a, 2014b), for example. The reasoning behind this is that people having a positive view of the market would already naturally hold long positions on the market. It is therefore rather when people change their opinion, say by becoming more negative about the market, or less bullish, that they will have the tendency to sell. Assuming it to be proportional to the percentage change in bullishness, *RB(t)*, as well as economic news, $\varphi(t)$, the return *r(t)* is given by

$$r(t) = \frac{RB(t)}{\mu} + \varphi(t), \; \mu > 0. \tag{8}$$

Here, $\varphi(t)$ is a Gaussian distributed variable with mean 0, describing a standard deviation that varies in time as a function of market sentiment:

$$\sigma(t) = \sigma_0 exp\left(\frac{RB(t)}{\beta}\right), \; \sigma_0 > 0, \quad \beta > 0. \tag{9}$$

The impact of market performance on decision-making can then be taken into account by letting $m_{k,j}(t)$ depend on the market performance according to

$$m_{k,j}(t) = m_{k,j}(t) exp\left(\frac{r(t)}{\alpha}\right), \; \alpha > 0, \; m_{kj} \leq 1. \tag{10}$$

In this way, the transition probabilities for a change of opinion, Label (9), depend directly on the market return over the last time period. The reasoning behind this dependence is that, if, for example, the market had a dramatic downturn at the close yesterday, then, in meetings the next morning, those with a bearish view will be more likely to convince even a bullish majority about their point of view.

Synchronization in the decision-making due to communication between people can now be studied via tipping point analysis, for example. Once extreme sentiment *B* = 0.1 has been created via synchronization, this can be used to identify a tipping point of the market: when, say, $B \to 1$, any further increase in *B* is limited, and this in turn limits further price increases in the market. However, any negative economic news, $\varphi(t)$, will then lead to a decrease in *B(t)* through Labels (7) and (9). The cases *B* = 0.1 therefore act as reflection points in the model, thereby making it possible to identify tipping points in the price dynamics of the markets. One illustration of such tipping point dynamics in real markets is shown in Figure 8, taken from (Andersen et al. 2014b). In (Andersen et al. 2014b), maximum likelihood methods were used to estimate the parameters of the model, after which an out-of-sample analysis was performed on the EUBanks index around the time of the financial crisis in 2008. As can be seen in Figure 8a,c, prior peaks, $B \cong 1$, in the estimated sentiment do indeed announce a tipping point in the performance of the index returns.

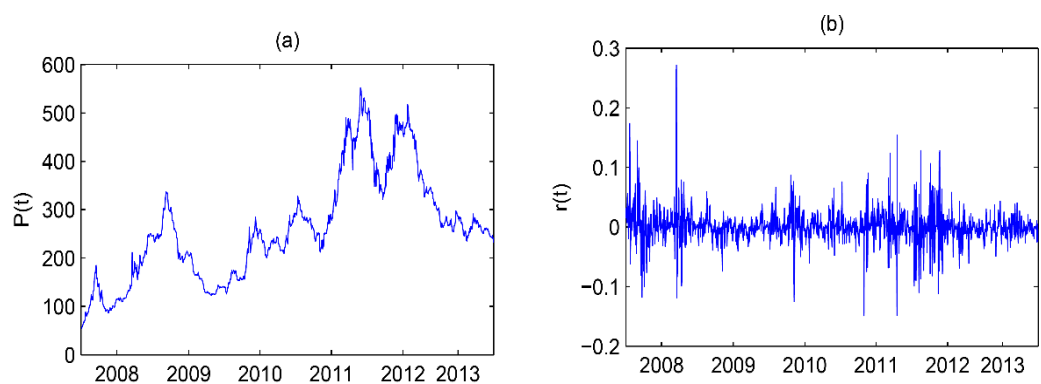

**Figure 8.** *Cont.*

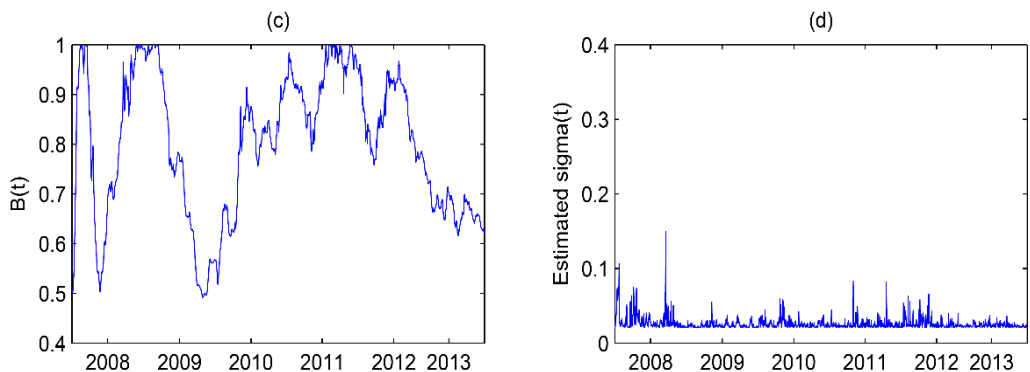

**Figure 8.** EUBanks index prices and returns, together with the corresponding estimated conditional volatilities and bullishness proportions under the assumption of a conditional Student-*t* distribution; (**a**) EUBanks index price *P*(*t*); (**b**) EUBanks index returns *r*(*t*); (**c**) estimated bullishness proportions *B*(*t*); (**d**) estimated conditional volatilities $\sigma(t)$, taken from (Andersen et al. 2014b).

## 3. Discussion

We have introduced three different models from the field of socio-finance to investigate three different pathways that could lead financial market participants to synchronize their decision-making, and thereby create a risk of contagious and volatile market phases. One pathway comes about when stock market indices, seen as a set of coupled integrate-and-fire oscillators, synchronize in frequency. Another occurs due to feedback mechanisms between market performance and the use of certain (decoupled) trading strategies. In addition, a third pathway could be brought about by communication and its impact on human decision-making.

Synchronization is a well-known phenomenon in economics, used to describe the way trading partners can introduce correlations in business cycles across international borders. The recent global financial crisis raises the question of the impact of financial market integration on synchronization of business cycles across borders. Another question is: Can synchronization created endogenously in financial markets spill over into the economy and thereby cause synchronization in business cycles across borders? It should be noted that very little research has been done on synchronization that is created endogenously by the financial markets themselves, without there necessarily being any economic cause. It is the authors' hope that the present article will fuel awareness of this topic.

**Author Contributions:** Conceptualization, J.V.A.; methodology J.V.A.; software N.M. and J.V.A.; validation N.M. and J.V.A; formal analysis N.M. and J.V.A.; investigation N.M. and J.V.A.; resources N.M. and J.V.A.; data curation N.M. and J.V.A.; writing—original draft preparation, J.V.A.; writing—review and editing, N.M. and J.V.A.; visualization, N.M. and J.V.A.; supervision J.V.A.; project administration J.V.A.: funding acquisition N.M. and J.V.A.

**Acknowledgments:** This work was carried out under the auspices of the Laboratory of Excellence on Financial Regulation (Labex ReFi), supported by PRES heSam under the reference ANR-10-LABX-0095. It benefited from the French government support, managed by the National Research Agency (ANR) as part of the project Investissements d'Avenir Paris Nouveaux Mondes (Paris-New Worlds investments for the future) under the reference ANR-11-IDEX-0006-02.

**Conflicts of Interest:** The authors declare no conflicts of interest.